\documentclass{JHEP3} 
\usepackage{amsmath}
\usepackage{epsfig}
\usepackage{amssymb,amsfonts,nicefrac}

\def\zi{\mathbb{Z}}

\def\di{\text{d}}

\def\slr{\text{SL(2},\mathbb{R})}
\def\slc{\text{SL(2},\mathbb{R})/\text{U}(1)}

\title{
\boldmath D-branes in Lorentzian AdS$_3$\footnote{Research partially supported by the EEC 
under the contract MRTN-CT-2004-512194.}
\unboldmath}
\author{Dan Isra\"el
\\
Racah Institute of Physics, The Hebrew University \\
Jerusalem 91904, Israel \\
E-mail:  \email{israeld@phys.huji.ac.il}
}

\abstract{
We study the exact construction of D-branes in Lorentzian AdS$_3$. We start by defining a family 
of conformal field theories that gives a natural Euclidean version of the $\slr$ CFT and does not 
correspond to H$_{3}^+$, the analytic continuation of AdS$_3$.   
We argue that one can recuperate the exact CFT results of Lorentzian AdS$_3$, upon 
an analytic continuation in the moduli space of these conformal field theories. 
Then we construct exact boundary states for various symmetric and symmetry-breaking 
D-branes in AdS$_3$.
}

\preprint{hep-th/0502159}

\begin{document}
\section{Introduction and motivation}
The three-dimensional anti-de-Sitter spacetime (AdS$_3$) has the unique virtue among other incarnations of the AdS /CFT 
holographic correspondence~\cite{Maldacena:1997re} of being solvable as a worldsheet theory. Indeed it corresponds to 
(the universal cover of) the \textsc{wzw} model for the group $\slr$. Moreover it is a good laboratory to 
probe thermodynamical issues of gravity, since its black holes solutions are simply quotients by 
a discrete group, the famous \textsc{btz} black holes~\cite{Banados:1992wn}. 

This exact conformal field theory is however difficult to solve, because the underlying group is 
non-compact. The correct spectrum 
itself has been found correctly only recently~\cite{Maldacena:2000hw} (inspired by~\cite{Henningson:1991jc}) 
and confirmed later by a computation of the partition function of the $\slr$ \textsc{wzw} model~\cite{Israel:2003ry}. 
Most of the other progress in this area has been made in the conformal field theory on the three-dimensional 
hyperbolic space H$_{3}^+$, i.e. the geometric coset $\mathrm{SL}(2,\mathbb{C}) / \mathrm{SU}(2)$, 
obtained by an analytic continuation of AdS$_3$. The correlators of this theory have been computed through hard work in 
a series of papers~\cite{Teschner:2001gi,Teschner:1999ug,Teschner:1997fv,Teschner:1997ft} using 
the chiral symmetries of the model. However it is quite difficult to extract the physics 
of the Lorentzian AdS$_3$ from this conformal field theory, which is non unitary because of the 
imaginary NS-NS two-form. In particular its spectrum contains neither discrete representations 
nor long strings, which are distinguishing features of the $\slr$ \textsc{wzw} model. Nevertheless 
the correlators of $\slr$ have been obtained in~\cite{Maldacena:2001km} by taking this difficult 
route. 

In this note, we argue that it is possible to give an alternative definition of Euclidean 
AdS$_3$, which is more natural from the point of view of conformal field theory. We 
will study string theory on the following family of backgrounds: 
\begin{eqnarray}
\di s^2 &=& k \left[ \di r^2 + \frac{R^2 \tanh^2 r \ \di \varphi^2 \ + \ \di x^2}{\tanh^2 r + R^2}
 \right]\nonumber \\
\mathcal{B} &=&  \frac{k\, \tanh^2 r}{\tanh^2 r + R^2} \ \di x \wedge \di \varphi \nonumber \\
e^{2\Phi} &=& \frac{R^2 \ e^{2\Phi_0}}{R^2 \, \cosh^2 r + \sinh^2 r}
\label{backEads}
\end{eqnarray}
parameterized by the modulus $R$. Note that the NS-NS two-form is real and the dilaton bounded from above 
everywhere. As we shall see the corresponding \textsc{cft} is 
unitary. This model captures most of the features of the $\slr$ conformal field theory; its spectrum contains 
the same discrete and continuous representations, and all sector of spectral flow as well, which translates 
here into combinations of winding modes. However the geometry is rather different from AdS$_3$ since this manifold has 
only a U(1) $\times$ U(1) group of isometries and asymptotes a linear dilaton background. The AdS$_3$ background 
itself (the metric and the NS-NS two-form) is obtained for the value $R^2 = -1$.

This family of theories is nothing but the the line of 
deformations of $\slr$ at level $k$ by the truly marginal operator $\delta S = \int d^2 z \ J^3 \bar{J}^3$ 
that has been studied in~\cite{Israel:2003ry}, with an imaginary parameter though. This is indeed 
an exact \textsc{cft}, T-dual to an orbifold of $\slc |_\textsc{a}  \times \text{U(1)}_{kR^2}$, where the first factor is 
the axially gauged \textsc{wzw} model, the well-known 
cigar \textsc{cft}~\cite{Elitzur:1991cb,Mandal:1991tz,Witten:1991yr}. The issue of the analytic continuation 
in this description of the theory is then trivialized since the would-be timelike direction is simply 
a free bosonic field, coupled to the $\slc$ model through its zero modes. This was (implicitly) 
the method used in~\cite{Israel:2003ry} to compute the partition function of the $\slr$ \textsc{wzw} 
theory. The prescription that we shall employ to compute AdS$_3$ quantities is to do all the 
worldsheet computations in the theory defined by eq.~(\ref{backEads}) --~which will be 
(analytic) functions of $R$~-- and at the end go back to the Minkowskian AdS$_3$ theory through 
the continuation $R \to i$. This defines an {\it analytic continuation in the moduli space} 
of the conformal field theory. One can easily 
check that the correlators computed in~\cite{Maldacena:2001km} can be obtained more simply 
using this prescription. Note that the analytic continuation in flat spacetime can be also 
formulated this way, and its extension to AdS$_3$ as we see is rather natural.

An essential ingredient to understand the physics of string theory in AdS$_3$ is to construct the 
D-branes in this spacetime. This is indeed the simplest setup to study D-branes in curved spacetimes 
and it has important implications for the holographic duality. These D-branes have been studied at the semi-classical level in~\cite{Bachas:2000fr,Petropoulos:2001qu,Lee:2001xe}. However the exact boundary states 
are not known.\footnote{see~\cite{Hikida:2001yi,Rajaraman:2001cr,Deliduman:2002bf} for previous works on this 
topic.}
The main purpose of this work is to construct them, extending the procedure described 
above for the closed string theory to the boundary \textsc{cft}. 
Indeed the D-branes in the H$_{3}^+$ theory has been constructed 
in~\cite{Ponsot:2001gt,Lee:2001gh}, and extended to the coset theory $\slc$ 
in~\cite{Ribault:2003ss} (see also~\cite{Israel:2004jt,Eguchi:2003ik,Fotopoulos:2004ut}). 
We will use these results to construct the D-branes boundary states of the Euclidean AdS$_3$ 
defined by eq.~(\ref{backEads}), and then in the Lorentzian AdS$_3$ 
spacetime itself. Another advantage of this method is that it allows to construct the symmetry-breaking 
D-branes (see~\cite{Maldacena:2001ky} for their analogues in the SU(2) \textsc{wzw} model) easily. 
Note finally that most of the expressions in this work will be valid not only for AdS$_3$ but also 
for the full line of $J^3 \bar{J}^3$ marginal deformations of $\slr$. In all the 
paper we will work in the bosonic string, although the extension to the supersymmetric case is rather simple. 

This paper is organized as follows. We begin in section~\ref{bulk} by defining precisely 
the bulk theory of Euclidean AdS$_3$. Then in the following sections we construct various 
class of D-branes. First we study in detail the much important AdS$_2$ D-branes in sect.~\ref{ads2bran}, 
as well as their symmetry-breaking counterparts. Then we move to the localized D-branes in sect.~\ref{localbranes}, 
and D-branes covering AdS$_3$ in section~\ref{extbranes}.
In the appendix we recall some facts about $\slr$ representations and characters 
that will be used extensively in this work.

\section{Euclidean AdS$_3$: closed string sector}
\label{bulk}
The N-th cover of the \textsc{wzw} model $\slr$ at level $k$ corresponds to a timelike compactification of  
AdS$_3$ spacetime, with a non-zero NS-NS electric flux:
\begin{eqnarray}
\di s^2 &=& k \left[ \di r^2 + \sinh^2 r \, \di \varphi^2 - N^2 \cosh^2 r \, \di x^2 \right]\\
\mathcal{B} &=& Nk \ \sinh^2 r \ \di \varphi \wedge \di x
\end{eqnarray}
where we have normalized the time coordinate such that $x \sim x + 2\pi$. The physically sensible case 
corresponds to the {\it infinite cover} of the group $\slr$ (i.e. $N \to \infty$), for which the globally 
defined time $x$ is non-compact. However it will be convenient in the following to keep $N$ 
arbitrary. 
After a T-duality along the time-like direction $x$ we obtain the following torsionless solution with a non-trivial 
dilaton:
\begin{eqnarray}
\di s^2 &=& k \left[ \di r^2 + \tanh^2 r \left( \di \varphi + \frac{\di x }{kN}\right)^2 
- \left( \frac{\di x}{kN} \right)^2 \right] \\
\Phi &=& \Phi_0 - \log \cosh r
\label{tdualgeom}
\end{eqnarray}
And after the redefinitions $\varphi + \nicefrac{x}{kN} = \phi$ and $t = \nicefrac{x}{kN}$ 
we get the conformal field theory $\slc \times U(1)$, but with the identifications 
$$(t,\phi)\sim (t + \nicefrac{2\pi n}{kN}, 
\phi+ \nicefrac{2\pi n}{kN}+2\pi m).$$ 
In other words this is the orbifold
\begin{equation}
\frac{\text{SL}(2,\mathbb{R})_k /\text{U(1)} \times \text{U(1)}_{-k}}{\zi_{Nk}}
\end{equation}
where $\text{U(1)}_{-k}$ means a compact time-like U(1) of radius $\sqrt{2k}$.\footnote{Here 
and in the following we work with $\alpha' = 2$.} the action of the orbifold is taken to be diagonal 
in the product \textsc{cft}. A very important point is, while for the universal cover this procedure is well defined 
for an arbitrary level $k$ (because we get a continuous orbifold), the theory can be properly 
defined for an arbitrary cover only if the level $k$ is integer (this can be generalized to rational $k$). 
In the following we will mostly deal with the case $k$ integer. 

In this T-dual definition of the AdS$_3$ string theory the analytic continuation to Euclidean 
space is rather trivial since the timelike direction is a free, compact boson coupled 
to the coset $\slc$ only through the action of the shift orbifold on its zero modes.
This can be generalized slightly by starting with the following T-dual model
\begin{eqnarray}
\di s^2 &=& k \left[ \di r^2 + \tanh^2 r \left( \di \varphi + \frac{\di x }{kN}\right)^2 
+ R^2 \left( \frac{\di x}{kN} \right)^2 \right] \nonumber \\
\Phi &=& \Phi_0 - \log \cosh r
\end{eqnarray}
with still the periodicity $x \sim x + 2\pi$. It corresponds simply to changing the radius 
of the now spacelike direction $x$. After T-dualizing back it gives the following solution:
\begin{eqnarray}
\di s^2 &=& k \left[ \di r^2 + \frac{R^2 \tanh^2 r \ \di \varphi^2 + N^2 \di x^2}{\tanh^2 r + R^2}
 \right]\\
\mathcal{B} &=&  \frac{Nk\, \tanh^2 r}{\tanh^2 r + R^2} \ \di x \wedge \di \varphi \nonumber \\
e^{2\Phi} &=& \frac{R^2 \ e^{2\Phi_0}}{R^2 \, \cosh^2 r + \sinh^2 r}
\end{eqnarray}
which is exactly the Euclidean AdS$_3$ space given in eq~(\ref{backEads}). 
In the limiting cases $R \to 0$ and $R \to \infty$ one obtains 
respectively the vector coset $\slc$ --~the trumpet~-- times a decoupled non-compact U(1), 
and the axial coset $\slc$ --~the cigar.\footnote{More precisely, starting 
with the $N$-th cover of AdS$_3$ this background interpolates between the 
$N$-th cover of the trumpet and the cigar (which is uniquely defined), 
see~\cite{Israel:2003ry} for details.} This solution can also be obtained 
directly by considering a coset $[\slr \times \text{U(1)}]/\text{U(1)}$, 
much as in~\cite{Giveon:1993ph} for SU(2). 
Note finally that this background (and its T-dual) is very 
much related to the background of NS5-branes of a circle 
studied in~\cite{Sfetsos:1998xd,Israel:2004ir} (see also~\cite{Ooguri:1995wj,Giveon:1999px}).\footnote{
The relation is indeed very accurate, since each section of the NS5-brane geometry for constant $\theta$ 
(see eqn.~(5.7) in~\cite{Israel:2004ir}) gives such a background with the identification 
$R=\text{cotan }\, \theta$.} As a consequence our study of D-branes will be very close to the 
analysis of D-branes in this NS5-brane background performed in~\cite{Israel:2005fn}.

To close this discussion, let us recall that our definition of Lorentzian AdS$_3$
will be given by the following analytic continuation:
\begin{equation}
\frac{\slc_k \times U(1)_{R^2 k}}{\zi_{Nk}}\quad \stackrel{R^2 \to -1}{\vector(1,0){50}}
\quad \slr_k
\end{equation}
and to get the physically desirable universal cover we take $N \to \infty$.
\subsection*{Closed string spectrum}
To construct the closed string spectrum of the theory, we start with the 
partition function of $\slc \times U(1)$ and implement the diagonal orbifold action 
in the standard way compatible with modular invariance. The partition 
function of the coset $\slc$ has been derived in~\cite{Hanany:2002ev} by means of a 
direct path integral approach (again, by using the H$_3^+$ results) 
and further analyzed in~\cite{Israel:2004ir}. It is built with the 
coset characters that are recalled in the appendix. First we have a spectrum of 
discrete representations, of real spin in the range 
$\nicefrac{1}{2} < j < \nicefrac{k-1}{2}$ given by:
\begin{equation}
Z_\text{dis} (\tau) = \sum_{n,w \in \zi} \int_{\frac{1}{2}}^{\frac{k-1}{2}} \di j \ 
\sum_{r,\bar{r} \in \zi} \ \delta(2j+r+\bar{r}+kw) \ \delta_{r-\bar{r},n} \ 
\lambda^{d}_{j,r} (\tau ) \ \bar{\lambda}^{d}_{j,\bar{r}} (\bar{\tau})
\end{equation}
in terms of coset characters $\lambda^{d}_{j,r} (\tau )$ of the discrete representations. 
The constraint enforced by the delta-function come from the coset construction. Note that this 
expression is valid for arbitrary $k$ (non-integer or integer). In the $k$ integer case, it can be recast 
in a more standard form since only half-integer values of the spin $j$ appear. Let us recall 
that the spectrum of primary states reads:
\begin{eqnarray}
L_0 &=& -\frac{j(j-1)}{k-2} + \frac{(n+kw)^2}{4k} \nonumber\\
\bar{L}_0 &=& -\frac{j(j-1)}{k-2} + \frac{(n-kw)^2}{4k}
\label{cosetspec}
\end{eqnarray}

The second part of the spectrum is made with continuous representations, of imaginary spin 
$j=\nicefrac{1}{2} + i P$. They correspond to asymptotic states that propagate in the cigar 
geometry, and consequently their spectrum is associated with an infinite volume divergence. 
This divergence can be regularized (much as in~\cite{Maldacena:2000kv}), and leads 
to the result:
\begin{equation}
Z_\text{cont} (\tau) = \int_{0}^\infty \di P \ \sum_{n,w \in \zi} \ 
\rho (P,n+kw,n-kw) \quad 
\lambda^{c}_{\nicefrac{1}{2}+i P,\frac{n+kw}{2}} (\tau ) \ 
\bar{\lambda}^{c}_{\nicefrac{1}{2}+i P,\frac{n-kw}{2}}  (\bar{\tau} ) \ 
\end{equation}
in terms of the characters $\lambda^{c}_{\nicefrac{1}{2}+i P,m}$ of the continuous representations. 
The density of states $\rho$ contains a leading part proportional to the 
(regularized) infinite volume of the manifold, and a subleading term given by 
the phase shift of the wave functions by the effective potential 
near the tip of the cigar~\cite{Hanany:2002ev}. The partition function contains 
also other non-universal contribution which appears because this infrared regulator 
breaks the chiral symmetries of the theory~\cite{Israel:2004ir}. 

The $U(1)_{R^2k}$ part of the theory is rather trivial, since it is simply a free boson. Its spectrum 
if given by the following partition function:
\begin{equation}
Z_\text{u(1)} = \frac{1}{\eta (\tau ) \bar{\eta} (\bar{\tau} )} \ 
\sum_{M,W \in \zi} \ q^{\frac{(M+kR^2 W)^2\!\!\!\!}{4kR^2}}\  \ \bar{q}^{\frac{(M-kR^2 W)^2 \!\!\!\!}{4kR^2}} 
\end{equation}
with $q = \exp 2i \pi \tau$. 

Let us move now to the orbifold theory that defines the Euclidean AdS$_3$ \textsc{cft}. 
In the closed string spectrum, the $\zi_{kN}$ orbifold has only an effect 
on the part of the $\slc$ spectrum corresponding to the momenta of the compact U(1) in the asymptotic cylinder 
geometry (see eq.~(\ref{cosetspec})), and 
the zero modes of the free U(1)$_{R^2k}$ theory. The modification on the left and right 
momentum modes is as follows:
\begin{eqnarray*}
\frac{\left(n + kw, n-kw \right)}{\sqrt{2k}} &\to& 
\frac{\left( n+kw -\frac{\gamma}{N},n-kw + \frac{\gamma}{N} \right)}{\sqrt{2k}} \ , \ \ 
\gamma \in \zi_{kN} \\
\frac{\left( M + R^2 k W , M-R^2 k W \right)}{R\sqrt{2k}} 
&\to& \frac{\left( n+kN p + R^2 k W + 
\frac{R^2   \gamma}{N}, n+kN\, p-R^2 k W - \frac{R^2 \gamma}{N} \right)}{R\sqrt{2k}}\ , \\
&& \hphantom{aaaaaaaaaaaaaaaaaaaaaaaaaaaaaaaaaaaaaa}  p \in \zi
\end{eqnarray*}
Indeed invariance under the orbifold action enforces the constraint $M=n \mod Nk$, 
and twisted sectors labeled by $\gamma \in \zi_{kN}$ are requested for the 
modular invariance of the one-loop torus amplitude. Taking $N \to \infty$ and the analytic 
continuation $R \to i$, one can check easily that one obtains the partition function 
of the $\slr$ \textsc{wzw} model computed in~\cite{Israel:2003ry}. Trading the discrete 
momentum $\nicefrac{\gamma}{kN}$ with the continuous parameter $s \in [0,1)$ in the $N \to \infty $ limit, 
the identification with the $\slr$ quantum numbers is as follows:
\begin{equation}
m = -\frac{n}{2} + \frac{k}{2}(W+s)  \ , \ \ 
\bar{m} = \frac{n}{2} + \frac{k}{2} (W+s) 
\label{idsl2qn}
\end{equation}
The combined winding number $W+w$ is identified with the sector of {\it spectral flow} 
of the $\widehat{\mathfrak{sl} (2,\mathbb{R} )}$ representation (see~\cite{Israel:2003ry} and the appendix),  
so the spectrum of the $\slr$ \textsc{cft} contains all the images under spectral flow  
of the discrete and continuous spectra. The 
spacetime energy of the state is given by $E=k(W+s)$, which can take any 
real value.
\subsection*{Bulk two-point function}
To fix the normalization of the vertex operators and compare with the open string case, we will 
now give the two-point function for closed strings vertex operators. We shall denote in the following 
the vertex operators of the orbifoldized theory as:
\begin{equation}
V^{j}_{n w \gamma p W} (z,\bar{z}) = \Phi^{\text{sl(2)/u(1)}}_{j,n,w-\frac{\gamma}{kN}} (z,\bar{z}) \quad 
\Phi^{\text{u(1)}}_{n+kN p,W+\frac{\gamma}{kN}} (z,\bar{z}). 
\label{defvertexops}
\end{equation}
Starting from the two-point function of the coset $\slc$ computed 
in~\cite{Giveon:1999px,Giveon:1999tq},  
the two-point function on the sphere is given as follows:
\begin{eqnarray}
\langle V^{j}_{n w \gamma p W} \ (z_1,\bar{z}_1) V^{j'}_{n' w' \gamma'  p' W'} (z_2,\bar{z}_2) \rangle  =  
| z_2 - z_1 |^{-4\Delta_{j n w \gamma p W}} 
\phantom{aaaaaaaaaaaaaaaaaaaaaaaaaa}
\nonumber \\ \times \ \delta_{n,-n'} 
\delta_{kN w-\gamma,-kNw'+\gamma'} \delta_{p,-p'} \delta_{W,-W'} 
 \left[ \delta (j+j' +1) + R\left( j, n,w-\frac{\gamma}{kN} \right) \delta (j-j') 
 \right]\nonumber\\
 \nonumber\\
  \text{with the reflection amplitude} \phantom{aaaaaaaaaaaaaaaaaaaaaaaaaaaaaaaaaa
  aaaaaa}
  \nonumber \\
 \nonumber \\
  R\left( j, n,w-\nicefrac{\gamma}{kN} \right) = \nu^{1-2j} \frac{
 \Gamma (1-2j) \Gamma (1+ \frac{1-2j}{k-2} )}{
 \Gamma (2j-1) \Gamma (1+ \frac{2j-1}{k-2})}
 \frac{\Gamma (j+\frac{n+kw-\nicefrac{\gamma}{N}}{2}) 
 \Gamma (j+\frac{n-kw+\nicefrac{\gamma}{N}}{2})}{
\Gamma (1-j+\frac{n+kw-\nicefrac{\gamma}{N}}{2}) 
\Gamma (1-j+\frac{n-kw+\nicefrac{\gamma}{N}}{2})}
 \nonumber \\
 \label{2ptfctbulk}
\end{eqnarray}
and the normalization constant $\nu = 
\Gamma (1-\nicefrac{1}{k-2})/ \Gamma (1+\nicefrac{1}{k-2})$. 
By taking the limit of the universal cover $N \to \infty$ and 
identifying the $\slr$ quantum numbers as~(\ref{idsl2qn}) 
we obtain the two-point function of Lorentzian AdS$_3$ given 
in~\cite{Maldacena:2001km}. This can be easily generalized to 
three-point functions. Note that the non-trivial part of the correlators is 
independent of the parameter $R$. This concludes our discussion of the bulk 
theory. We now move to the boundary theory in the following 
sections.

\section{AdS$_2$ D-branes}
\label{ads2bran}
We start the discussion of the boundary conformal field theory 
with the construction of AdS$_2$ branes in the Lorentzian AdS$_3$.
These D-branes were found to be associated with {\it twined conjugacy classes} 
of $\slr$~\cite{Bachas:2000fr}. They are certainly the more interesting 
D-branes to consider since they have a well defined holographic 
interpretation; they correspond to domain walls in the spacetime \textsc{cft} 
dual to AdS$_3$~\cite{Bachas:2001vj}. The boundary states for the associated D-branes in H$_{3}^+$ have
been found in~\cite{Ponsot:2001gt,Lee:2001gh} by solving factorization constraints involving degenerate 
operators. Our aim is to construct them in 
Lorentzian AdS$_3$, by starting with the Euclidean AdS$_3$ \textsc{cft} defined above.  We will  
start with the appropriate D-branes in $\slc \times U(1)$ and move to the orbifold theory.

The AdS$_2$ D-brane of $\slr$ descends in the coset theory $\slc$ of the cigar to infinite 
D1-branes of embedding equation: 
\begin{equation}
\sinh r_0 = \sinh r \sin \left( \varphi  -\varphi_0\right)
\end{equation}
The parameter $\varphi_0$ corresponds to the position of this D1-brane on the compact circle at 
infinity (since the cigar asymptotes $\mathbb{R}_\mathcal{Q} \times S^1$) and the parameter $r_0$ gives 
the distance of the D-brane to the tip of the cigar $r=0$. 
The exact one-point function on the disk corresponding to this D1-brane has been obtained from descent of 
H$_{3}^+$ as~\cite{Ponsot:2001gt,Ribault:2003ss}:\footnote{Here and in the following we have
  suppressed the $(z,\bar{z})$ dependence. One should read 
$\langle V_{\alpha,\bar{\alpha}} (z,\bar{z}) \rangle_{\hat{\beta}} =
  |z-\bar{z}|^{-\Delta_{\alpha}-\Delta_{\bar{\alpha}}} \Psi_{\hat{\beta}} (\alpha,\bar{\alpha}) 
$ and the coefficient 
$\Psi_{\hat{\beta}} (\alpha,\bar{\alpha})$, including the selection rules, is given in the text.} 
\begin{eqnarray} 
\langle \Phi^{\text{sl(2)/u(1)}}_{j,nw} \rangle_{r_0,\varphi_0}
= \delta_{w,0} \frac{1}{\sqrt{2}} \left(\frac{k-2}{k}\right)^{1/4} \  \nu^{\nicefrac{1}{2}-j} \ 
\ e^{i n \varphi_0}\phantom{aaaaaaaaaaaaa} \nonumber \\ 
\left[ e^{-r_0(1-2j)}+(-1)^n e^{r_0(1-2j)}\right] 
\frac{\Gamma(1-2j) \Gamma(1+\frac{1-2j}{k-2})}{\Gamma(1-j+\frac{n}{2})\Gamma(1-j-
\frac{n}{2})}
\end{eqnarray}
This expression contains only coupling of closed string modes belonging to the continuous representations, 
for $j = \nicefrac{1}{2} + i P$. There are no poles associated to discrete representations 
of $\slr$ that would signal the coupling of the D-brane to the corresponding closed string states. 
However the gamma-functions in the numerator give poles associated to the infinite volume of 
the target space (they are "bulk poles" using the classification of~\cite{Aharony:2004xn}). 

To obtain a AdS$_2$ D-brane we have to choose for the extra $U(1)_{R^2 k}$ 
theory Dirichlet boundary conditions 
(otherwise we obtain the symmetry-breaking D-branes constructed at the end of this section). The 
corresponding one-point function is simply:
\begin{equation}
\langle \Phi^{\text{u(1)}}_{M,W} \rangle_{x_0} = \delta_{W,0} \ \frac{e^{i M x_0}}{(\sqrt{2k} R)^{1/2}}  
\end{equation}
for a D-brane located at $x=x_0$. 
Now let us move to the (Euclidean) AdS$_3$ \textsc{cft}.

\subsection*{Boundary states for AdS$_2$ branes in AdS$_3$}
In the geometry~(\ref{tdualgeom}) of the orbifold theory we start with a D-brane of embedding equation:
\begin{equation}
\sinh r_0 = \sinh r \sin \left( \varphi + \frac{x}{kN} -\varphi_0\right) \ , \ \ 
\frac{x}{kN}=x_0
\end{equation}
corresponding to a D1-brane of the cigar and a D0-brane of $U(1)$ as discussed above.
After a T-duality along $x$ we obtain the profile of the D-brane upon 
eliminating the dualized coordinate $x$, which is then identified with the gauge field. 
This general technique is discussed in~\cite{Simon:1998az} and has been heavily 
used in~\cite{Israel:2005fn} for the case of D-branes in NS5-geometries, which is 
very much related to  our problem as we already discussed. It gives 
a D2-brane in (Euclidean) AdS$_3$ of profile:
\begin{eqnarray}
\sinh r_0 &=& \sinh r \sin (\varphi + x_0 - \varphi_0). 
\end{eqnarray}
The parameter $x_0$ can be absorbed in a redefinition 
of the (continuous) parameter $\varphi_0$. 
It is interesting to remark that the embedding equation and the gauge field are the same for all the geometries~(\ref{backEads}), 
irrespectively of the actual value of the parameter $R$. In the Lorentzian AdS$_3$ geometry ($R^2=-1$), 
the embedding equation of the D-brane defines an AdS$_2$ submanifold of AdS$_3$.

Note that the gauge for the B-field is fixed implicitly by the particular 
T-duality we use for the construction of the D-branes. In this gauge the magnetic field on the D-branes vanishes. However, 
we can choose instead the gauge 
\begin{equation} 
\mathcal{B}' = Nk \ \cosh^2 r \ \di x \wedge \di \varphi
\end{equation} for the NS-NS two-form in AdS$_3$. Then the T-dual background 
will be given by \begin{equation} \di s^2 = k \left[ \di r^2 + \tanh^2 r \left( \frac{\di x}{kN}\right)^2 - 
\left( \di \varphi - \frac{\di x}{kN}\right)^2 \right]\, , \end{equation}
giving by the same reasoning the following magnetic field on the D-brane:
\begin{equation}
F = \frac{kN}{2\pi}\  \di  \varphi \wedge \di x = - \frac{kN}{2\pi} \ \frac{\sinh r_0 \coth r}{\sqrt{\sinh^2 r - \sinh^2 r_0}} \ 
\di r \wedge \di x \, ,
\end{equation}
the two expressions being related by the embedding equation. 

Let us now consider the exact \textsc{cft} description of those D-branes. 
To construct  the D-branes in the $\zi_{kN}$ diagonal orbifold \textsc{cft} 
we have to sum over the images 
$(\varphi_0 +  \nicefrac{2\pi \ell}{kN} , x_0 -  \nicefrac{2\pi \ell}{kN} )$, 
$\ell \in \zi_{kN}$. It will impose $M=n \mod kN$ which is the condition of invariance 
of the closed string state under the orbifold action. Using the labeling 
of the primary states given by eq~(\ref{defvertexops}), we obtain the 
following one-point function for an AdS$_2$ D-brane of 
parameters $(r_0,\varphi_0,x_0)$:  
\begin{eqnarray}
\langle  V^{j}_{n w \gamma p  W } \rangle_{r,\varphi_0} 
&=& 
\delta_{Nk\, w-\gamma,0} \ \delta_{Nk\, W+\gamma,0}\  \nu^{\nicefrac{1}{2}-j} \left( 
\frac{k-2}{2} \right)^{1/4} \frac{1}{\sqrt{2R}} \ 
e^{i n \,(\varphi_0+x_0 )} \  e^{i kNp\, x_0} \nonumber \\
&& \left[ e^{-r_0(1-2j)}+(-1)^n e^{r_0(1-2j)} \right]
\frac{\Gamma(1-2j) \Gamma(1+\frac{1-2j}{k-2})}{\Gamma(1-j+\frac{n}{2})\Gamma(1-j-
\frac{n}{2})}\nonumber \\
\label{oneptads2}
\end{eqnarray}
This expression is valid for all the family of Euclidean AdS$_3$ solutions of 
eq.~(\ref{backEads}). It gives also the boundary state for the AdS$_2$ D-brane in 
Lorentzian AdS$_3$  by taking the analytic continuation $R \to i$. 

\subsection*{Annulus amplitude}
To compute the spectrum of open strings (and check the consistency of this D-brane), 
we will use the channel duality of the annulus amplitude~\cite{Cardy:1989ir}. 
We start by writing the annulus amplitude in the closed string channel, which is obtained 
from the closed string spectrum of continuous representations in the orbifold theory 
and the overlap of the boundary states whose coefficients are given by 
the one-point function~(\ref{oneptads2}). Since this boundary states contains only 
Ishibashi states associated to continuous representations (as in $\slc$) we obtain the 
following amplitude between two D-branes of parameters $(r_0,\varphi_0,x_0)$ 
and $(r_0 ',\varphi_0 ',x_0 ')$, with $\tilde{\tau}=-1/\tau$:
\begin{eqnarray}
Z^{\textsc{cl}} (\tilde{\tau}) = \frac{1}{2\sqrt{2(k-2)}R} 
  \sum_{n,p \in \zi} e^{i (n+kNp) (x_0 ' - x_0 )}
\frac{\tilde{q}^{\frac{(n+kNp)^2}{4k R^2}}}{\eta (\tilde \tau )} 
\int_{0}^{\infty} \di P \ \frac{\cosh 2\pi P + (-)^n}{\sinh 2\pi P \sinh \frac{2\pi P}{k-2}}
\nonumber \\ e^{i (\varphi_{0}' - \varphi_0 ) n} \ 
\left\{ \cos 2P (r_0 - r_0 ') + (-)^n \cos  2P (r_0 + r_0 ' ) \vphantom{\frac{1^1}{1_1}} \right\} 
\lambda^{c}_{\nicefrac{1}{2}+i P, \frac{n}{2}} (\tilde{\tau} )\nonumber \\
\end{eqnarray}
Then we can modular transform to the open string channel, using the formulas discussed in the appendix.
There is a divergence associated 
to the infinite volume available to the open strings. To remove this universal divergence 
from the amplitude we construct 
the {\it relative partition function} w.r.t. to a spectrum of open string stretched between 
reference D-branes of parameters $(r_\ast,\varphi_0,x_0)$ and 
$(r_\ast ',\varphi_0 ',x_0 ')$, see~\cite{Ponsot:2001gt}. 
We obtain the following one-loop amplitude for the open strings stretched between to AdS$_2$ branes 
in Euclidean AdS$_3$:
\begin{eqnarray}
Z^\textsc{op}_{(r_0,\varphi_0,x_0;r_0 ',\varphi_0 ',x_0 ')} (\tau) 
- Z^\textsc{op}_{(r_\ast,\varphi_0,x_0; r_\ast ' ,\varphi_0 ' ,x_0 ' )} (\tau)
 = \frac{1}{N}
\sum_{m,w \in \zi} \frac{1}{\eta (\tau) } \ q^{\frac{R^2}{k} 
\left( \frac{m}{N} + \frac{k(x_0 ' - x_0 )}{2\pi} \right)^2}
\nonumber \\ \sum_{W \in \zi} \ \int \di P' \ \left\{ \frac{\partial
\log \frac{R (P'|r_0 ,r_0 ')}{R (P'|r_{\ast},r_{\ast}')} }{2i\pi \partial
P'}\,
\lambda^{c}_{P',-\frac{m}{N}+kW + \frac{k(\varphi_0 ' - \varphi_0 )}{2\pi}} (\tau )
 \right. \phantom{aaaaaaaaaaaa} \nonumber \\
+ \left. \frac{\partial
\log \frac{R (P'|r_0 ,-r_0 ')}{R (P'|r_{\ast},-r_{\ast}')} }{2i\pi\partial P'}\,
\lambda^{c}_{P',-\frac{m}{N}+k(W+\nicefrac{1}{2}) + \frac{k(\varphi_0 ' - \varphi_0 )}{2\pi}} (\tau ) 
\right\} 
\nonumber \\
\label{openspecads2}
\end{eqnarray}
The first and second term contains integer and half-integer windings respectively. 
Note that the normalization of the open string partition function has be chosen to be compatible with 
the limit of the infinite cover (continuous orbifold), see below. 
This result is expressed in terms of the reflection amplitude for open strings 
streched two AdS$_2$ D-branes $(r,r')$ in H$_3^+$, 
whose $(r,r')$-dependent part reads~\cite{Ponsot:2001gt,Ribault:2002ti}:
\begin{equation}
R (P|r,r') = \frac{S_{k}^{(0)} \left(\frac{k-2}{2\pi} (r+r') +P \right)}{
S_{k}^{(0)} \left(\frac{k-2}{2\pi} (r+r') -P \right)}
\frac{S_{k}^{(1)} \left(\frac{k-2}{2\pi} (r-r') +P \right)}{
S_{k}^{(1)} \left(\frac{k-2}{2\pi} (r-r') -P \right)}
\, ,
\end{equation}
in terms of the special functions:
\begin{eqnarray}
\log S^{(0)}_k (x) &=& i \int_{0}^{\infty} \frac{dt}{t}
\left( \frac{\sin \frac{2tx}{k-2}}{2\sinh t\,  \sinh \frac{t}{k-2}} - \frac{x}{t} \right)
\\
\log S^{(1)}_k (x) &=& i \int_{0}^{\infty} \frac{dt}{t}
\left( \frac{\cosh t\, \sin \frac{2tx}{k-2}}{2\sinh t\,  \sinh \frac{t}{k-2}} - \frac{x}{t}
\right)
\end{eqnarray}
We discuss this reflection amplitude in more detail now.
\subsection*{The boundary two-point function}
As in the bulk the reflection amplitude comes from the two-point function of boundary operators:
\begin{equation}
\langle \Phi^{rr'}_{j,\mu} (x)\Phi^{rr'}_{j' ,-\mu} (y) \rangle = | x-y|^{-2\Delta_{j,\mu}} \delta (j-j') \ R(j,\mu | r,r')
\end{equation}
with $$\mu = -\frac{m}{N}+kW + \frac{k(\varphi_0 ' - \varphi_0 )}{2\pi}$$ or 
$$\mu = -\frac{m}{N}+k(W+\nicefrac{1}{2}) + \frac{k(\varphi_0 ' - \varphi_0 )}{2\pi}.$$ 
The full reflection amplitude for the AdS$_2$ D-branes in H$_3^+$ has been computed 
in~\cite{Ponsot:2001gt} for open strings with both ends on the same D-brane 
and in~\cite{Ribault:2002ti} in the generic case. We can then obtain from descent the 
corresponding reflection amplitude in the coset $\slc$ (by means a Fourier transform much as 
in~\cite{Giveon:1999px} and explained in the appendix) and then lift the result to the Euclidean AdS$_3$ by means of the 
orbifold construction. Then the full expression for the reflection amplitude reads:
\begin{eqnarray}
R(j,\mu | r,r') =  \nu^{\nicefrac{1}{2}-j} \ \frac{\Gamma^{2}_k (k-2+j)}{\Gamma^{2}_k (k-2+1-j)}
\frac{\Gamma_k ( k-2 +1-2j)}{\Gamma_k ( k-2 +2j-1)} \nonumber\\
\frac{S_{k}^{(0)} \left(\frac{k-2}{2\pi} (r+r') -i(j-1/2) \right)}{
S_{k}^{(0)} \left(\frac{k-2}{2\pi} (r+r') +i(j-1/2) \right)}
\frac{S_{k}^{(1)} \left(\frac{k-2}{2\pi} (r-r') -i(j-1/2) \right)}{
S_{k}^{(1)} \left(\frac{k-2}{2\pi} (r-r') +i(j-1/2) \right)}
\frac{S_{k}^{(0)}(+i(j-1/2))}{S_{k}^{(0)}(-i(j-1/2))} \nonumber \\
\frac{\Gamma (j-\mu) 
}{\Gamma (1-j-\mu)}
\nonumber \\
\label{boundreflampl}
\end{eqnarray}
where the last factor comes from the Fourier transform, see appendix. This expression involves the non-trivial 
function
\begin{equation}
\Gamma_k (x) = (k-2)^{-\frac{x(x+2-k)}{2(k-2)}} (2\pi)^{x/2} \Gamma_{2}^{-1} 
(x|1,k-2)
\end{equation}
written in terms of $\Gamma_2 (x|\omega_1,\omega_2)$ the Barnes double gamma-function.

This expression has indeed poles signaling the presence of discrete representations, for 
$\mu-j \in \mathbb{N}$. Note that for the coset $\slc$ ($\mu = kW$) 
there are also poles of this kind provided $k$ is non-integer.\footnote{ 
Indeed, using the range allowed for $j$, we find that the denominator is negative, and integer for $k$ integer.} 
This seems to contradict the fact that no discrete representations appear in the open string 
spectrum~(\ref{openspecads2}), and the related observation that the one-point function~(\ref{oneptads2}) does not 
contains poles corresponding to couplings of closed string modes in the discrete representations with 
the D-brane. To solve this puzzle, one can invoke the fact 
(this was suggested in~\cite{Ribault:2003ss} for the corresponding brane in the coset) that what we really compute is a relative 
partition function, and the open string spectrum is expected not to depend on the parameters $r_0$ and 
$r_0 '$ of the D-branes. Indeed semi-classical analysis~\cite{Petropoulos:2001qu} and the computation of the 
free energy~\cite{Lee:2001gh} suggest that the open string spectrum contains a full discrete spectrum with all 
allowed values in the range $\nicefrac{1}{2} < j < \nicefrac{k-1}{2}$~-- much as the closed string spectrum.\footnote{
I thank Jan Troost for discussions on this point.}

\subsection*{Open string spectrum in Lorentzian spacetime}
Now let us go discuss the open string spectrum on the AdS$_2$ D-branes 
in Lorentzian AdS$_3$. As we already discussed in detail we have first to take the limit $N \to \infty$ 
in~(\ref{openspecads2}) to go to the universal cover of 
the $\slr$ group manifold. We have to replace the sum $\frac{1}{N} \sum_m f(m/N)$ by the Riemann integral 
$\int dE \ f(E)$ (and we can absorb $x_0$ by a shift of $\varphi_0$).
Then we can make the analytic continuation $R \to i$ to go back to Lorentzian AdS$_3$. To express the result 
in terms of $\slr$ characters we have to choose $\varphi_0 ' = \varphi_0 \mod 2\pi$. This is indeed 
necessary in order that the D-brane configuration preserve an $\slr$ symmetry.
Then we find the following open string partition function on the AdS$_2$ branes:
\begin{eqnarray}
Z_{\text{AdS}_2}^\textsc{op} (\tau ) = \int d E \ 
\sum_{w \in \zi} \int \di P' \ \left\{ \frac{\partial
\log \frac{R (P'|r_0 ,r_0 ')}{R (P'|r_{\ast},r_{\ast}')} }{2i\pi \partial
P'}\,
\chi^{c \, (2w)}_{\nicefrac{1}{2} + i P',E } \ (\tau ) \right. \phantom{aaaaaaaaaaaa} \nonumber \\
+ \left. \frac{\partial
\log \frac{R (P'|r_0 ,-r_0 ')}{R (P'|r_{\ast},-r_{\ast}')} }{2i\pi\partial
P'}\  
\chi^{c \, (2w+1)}_{\nicefrac{1}{2} + i P',E } \ (\tau )
\right\} 
\nonumber \\
\end{eqnarray}
written in terms of the $\slr$ characters $\chi^{c \, (w)}_{\nicefrac{1}{2} + i P',m}$ in the 
$w$-flowed representation. They are obtained from the coset characters $\lambda^{c}_m$ and the U(1) characters 
with the branching relations given in the appendix. This gives the following open string spectrum 
\begin{equation}
L_0 = \frac{P^2+\nicefrac{1}{4}}{k-2} - 2(w+\epsilon/2)\, E + 
k (w+\epsilon/2)^2 \ , \ \ \epsilon=0,1
\end{equation}
This is exactly what we expect for continuous representations with $m=E$ and $(2w+\epsilon)$ units of spectral flow. 
The open strings with even spectral flow have both ends at the same point on the D-brane, and 
those with odd spectral flow end on different points (they 
correspond to half-integer windings). Thus the on-shell continuous spectrum 
on the AdS$_2$ D-branes consists in long strings, in agreement 
with the semi-classical analysis of~\cite{Lee:2001xe}.

\subsection*{Symmetry-breaking D-brane}
As in SU(2)~\cite{Maldacena:2001ky} we can construct "B-branes" preserving only a 
$\widehat{\mathfrak{u}(1)}$ affine subalgebra of $\widehat{\mathfrak{sl} (2,\mathbb{R})}$. 
We already stressed in the introduction that the decomposition of $\slr$ as $\slc \times$U(1) is especially 
convenient to construct those D-branes. 

The first category of such D-branes are the symmetry-breaking cousins of the AdS$_2$ D-branes. 
To obtain them we combine a D1-brane of the axial coset 
$\slc$ (the cigar) with a Neumann D-brane of U(1)$_{R^2 k}$. 
Going through the same steps of T-duality, 
we find that these symmetry breaking D2-branes cover 
the region $r>r_0$ of AdS$_3$ and carry the following electric field, for instance in the first gauge choice  
(see also~\cite{Sarkissian:2002ie,Quella:2002ns}):
\begin{equation}
F = -\frac{kN}{2\pi} \left[ \frac{\sinh r_0 \tanh r}{\sqrt{\sinh^2 r - \sinh^2 r_0}} \ \di r \wedge \di x + \di \phi \wedge \di x 
\right]
\end{equation}

Using a similar reasoning as for the AdS$_2$ branes,  we find the following one-point function 
for the symmetry-breaking D-brane:
\begin{eqnarray}
\langle V^{j}_{n w \gamma p  W }  \rangle_{r_0} 
&=& 
\delta_{Nk\, w+\gamma,0} \ \delta_{n+kNp,0}\  \sqrt{\frac{R}{2}} \left( \frac{k-2}{2} \right)^{1/4} \ \nu^{1/2-j}  
 \   \nonumber \\
&& \left[ e^{-r_0(1-2j)}+ e^{i \pi kNp}\ e^{r_0(1-2j)} \right] 
\frac{\Gamma(1-2j) \Gamma(1+\frac{1-2j}{k-2})}{\Gamma(1-j+\frac{kNp}{2})\Gamma(1-j-\frac{kNp}{2})   }\nonumber \\
\end{eqnarray}
The only difference with the symmetric D-branes is that the boundary conditions along the two U(1) directions 
(the compact U(1) of the cigar and the extra U(1)$_{R^2 k}$) are different; 
indeed the annulus amplitude in the closed string channel will involve only U(1) characters  
with no momentum. 
Thus after a modular transform we get the following open string spectrum 
(for the universal cover of the Lorentzian AdS$_3$):\footnote{Of course the two terms can be 
brought together but this is not the case for an arbitrary cover.}
\begin{eqnarray}
Z^\textsc{op} (\tau ) = \int \di E \ 
 \frac{1}{\eta (\tau) } \  q^{-\frac{E^2}{k}} 
\nonumber \\ \int_{0}^{\infty} \di P' \ \int d\mu \ \left\{ \frac{\partial
\log \frac{R (P'|r_0 ,r_0 ')}{R (P'|r_{\ast},r_{\ast}')} }{2i\pi \partial
P'}\ \lambda^{c}_{ \nicefrac{1}{2} + i P',\mu} (\tau )
 + \frac{\partial
\log \frac{R (P'|r_0 ,-r_0 ')}{R (P'|r_{\ast},-r_{\ast}')} }{2i\pi\partial
P'}\ \lambda^{c}_{ \nicefrac{1}{2} + i P',\mu +\nicefrac{k}{2}} (\tau )
\right\} 
\nonumber \\
\end{eqnarray}
Since this partition function cannot be expressed in terms of $\widehat{\mathfrak{sl} (2,\mathbb{R})}$
characters, it is clear that this D-brane breaks the $\widehat{\mathfrak{sl} (2,\mathbb{R} )}$ 
symmetry of the \textsc{cft}.


\section{Localized D-branes}
\label{localbranes}
We now move to D-branes that are localized in space, at the origin $r=0$ of global coordinates in AdS$_3$. 
In the H$_3^+$ \textsc{cft} there is a series of D-branes corresponding to spheres of imaginary radius, 
labeled by an integer label $u=1,2,\cdots$. 
Even if their geometrical interpretation is not clear, they can be reduced to 
D-branes in the cigar \textsc{cft}, corresponding to D0-branes sitting 
at the tip $r=0$~\cite{Ribault:2003ss}. However their open string 
spectrum is associated to a {\it finite} representation of $\widehat{\mathfrak{sl} (2,\mathbb{R} )}$  of 
spin $j=-(u-1)/2$, which is generically not unitary~\cite{Israel:2004jt}. 
Thus only the D-brane with $u=1$ is physical because it corresponds to 
a trivial representation which is indeed unitary. Note also that 
this particular D-brane has a clear geometrical interpretation in H$_3^+$ as 
a point-like object. 
The associated one-point function for the D0-brane in the cigar is given by~\cite{Ribault:2003ss}:
\begin{equation}
\langle \Phi_{j\, nw} \rangle = \delta_{n,0} \  \left( \frac{k}{k-2}\right)^{1/4}\ 
\frac{\nu^{\nicefrac{1}{2}-j}}{\sqrt{k-2}} (-)^w
\frac{\Gamma (j+\frac{kw}{2}) \Gamma (j-\frac{kw}{2})}{\Gamma (2j-1) \Gamma (1+ \frac{2j-1}{k-2})}
\end{equation}
This has been generalized in~\cite{Israel:2004jt} to D0-branes in an orbifold that we will need afterwards.
\subsection*{Symmetric D-brane: D-instanton}
The symmetric D-brane associated to the D0-brane in the cigar is a D-instanton 
in AdS$_3$~\cite{Bachas:2000fr}. It is obtained 
by starting with the D0-brane of the cigar together with a Neumann D-brane for the extra U(1). 
More precisely, for the single cover of $\slr$ we obtain two copies of the D-instanton 
(for $x=0$ and $x =\pi$) associated to the center of $\slr$ 
(see e.g.~\cite{Quella:2002ns}), and for the N-th cover we have N copies of this picture, 
therefore N copies of the open string spectrum. 
Thus we will restrict our discussion to the single cover.

Following the same logic as before, we obtain the following one-point function for the D-instanton in AdS$_3$:
\begin{equation}
\langle V^{j}_{n w \gamma p  W}  \rangle = \frac{\sqrt{R}\  k}{2^{1/4}(k-2)^{3/4}}  \ 
\delta_{n,0} \ \delta_{p,0} \ 
 \ \nu^{\nicefrac{1}{2}-j}
\frac{\Gamma (j+\frac{kw-\gamma}{2}) \Gamma (j-\frac{kw-\gamma}{2})}{
\Gamma (2j-1) \Gamma (1+ \frac{2j-1}{k-2})}
\label{oneptdm1}
\end{equation}
This one-point function contains both couplings to the closed string states of the continuous representations 
($j = \nicefrac{1}{2} + i P$) and of the discrete representations ($\nicefrac{1}{2} < j < \nicefrac{k-1}{2}$). 
To compute the contribution of the latter to the closed string annulus amplitude, we take simply the residue 
at the corresponding simple pole. 
Here for simplicity we will only give the continuous part of the closed string annulus amplitude 
(more details can be found in~\cite{Ribault:2003ss,Israel:2004jt}):
\begin{eqnarray}
Z^\textsc{cl} (\tilde{\tau} ) = \frac{k^2 R}{\sqrt{2(k-2)}} \sum_{w,W \in \zi} \sum_{\gamma \in \zi_k} 
\int_{0}^\infty dP \  
\frac{\sinh 2\pi P \ \sinh \frac{2\pi P}{k-2}}{\cosh 2\pi P + \cos \pi (k w-\gamma)}
\nonumber \\ 
\frac{\tilde{q}^{\frac{k R^2}{4} (W+\gamma)^2}}{\eta (\tilde \tau )} \ \lambda^{c}_{\nicefrac{1}{2}+i P,\,  
k w-\gamma} (\tilde{\tau})
\end{eqnarray}
The modular transformation from the open string channel to the closed string channel, giving the  
closed string channel annulus amplitude of the D0-brane 
(for both continuous and discrete states) in $\slc$ is non-trivial and has been 
computed in~\cite{Ribault:2003ss,Israel:2004jt}. In our case the boundary state defined by~(\ref{oneptdm1}) corresponds to 
the following open string partition function for the D(-1)-brane of Euclidean AdS$_3$:
\begin{eqnarray}
Z^\textsc{op}_{D(-1)} (\tau ) = \sum_{r,w \in \zi} \lambda^{\mathbb{I}}_{r}
 (\tau) \ 
\frac{q^{\frac{(r+kw)^2}{kR^2}}}{\eta (\tau)} 
\end{eqnarray}
such that for  Lorentzian AdS$_3$ we get simply the identity representation of $\slr$ and its 
images under spectral flow, see the appendix.

\subsection*{Symmetry-breaking D-brane: D-particle}
The symmetry breaking D-branes associated to these localized branes are D-particles sitting 
at the origin $r=0$. In this case one take a Dirichlet D-brane for the extra U(1). 
Then we obtain the following one-point function:
\begin{equation}
\langle V^{j}_{n w \gamma p  W} \rangle = \   [2(k-2)]^{-1/4} \sqrt{\frac{k}{R(k-2)}} \ 
\delta_{n,0} \ \delta_{kNW+\gamma,0} \ e^{i x_0 \, kN p}
 \  \nu^{\nicefrac{1}{2}-j}
\frac{\Gamma (j+\frac{kw}{2}) \Gamma (j-\frac{kw}{2})}{
\Gamma (2j-1) \Gamma (1+ \frac{2j-1}{k-2})}
\end{equation}
It gives the following closed strings amplitude for the continuous representations:
\begin{eqnarray}
Z^\textsc{cl}  (\tilde{\tau}) =  \frac{k}{R\sqrt{2(k-2)}}\sum_{w,p \in \zi} \int_{0}^{\infty} dP \  e^{i (x_0 ' - x_0) kN p}
\phantom{aaaaaaaaaaaaaaaaa}\nonumber \\ \phantom{aaaaaaaaaaaaaaaaaaaaaaaa}
\frac{2\sinh 2\pi P \ \sinh \frac{2\pi P}{k-2}}{\cosh 2\pi P + \cos \pi k w}  
\ \frac{\tilde{q}^{\frac{kN^2}{4R^2 }p^2}}{\eta (\tilde{\tau}) }\  \lambda^{c}_{\nicefrac{1}{2}+iP, k w} 
(\tilde{\tau})
\end{eqnarray}
with again a contribution of discrete representations that can be found in~\cite{Ribault:2003ss,Israel:2004jt} 
for the coset $\slc$. The computation is exactly similar since for these symmetry-breaking D-branes the 
cigar \textsc{cft} and the extra $U(1)$ \textsc{cft} are decoupled.
This annulus amplitude in the closed string channel is compatible with the following open string partition function:
\begin{equation}
Z^\textsc{op}_{D0} (\tau )  = \frac{1}{N} \sum_{r \in \zi} \lambda^\mathbb{I}_{r} ( \tau) \sum_{m \in \zi} 
\frac{q^{\frac{R^2}{k} \left( \frac{m}{N}-\frac{k(x_0 ' -x_0)}{2\pi} \right)^2 }}{\eta (\tau)}
\end{equation}
written in terms of the characters $\lambda^\mathbb{I}_{r}$ of the identity representation for the coset.  
In the universal cover of Lorentzian AdS$_3$ we have simply
\begin{equation}
Z^\textsc{op}_{D0} (\tau)  = \sum_{r} \lambda^\mathbb{I}_{r} ( \tau) 
\int dE \  \frac{q^{-\frac{E^2}{k}}}{\eta (\tau)}
\end{equation}
and the spectrum of zero modes is the same as for open strings attached to a D-particle in flat space.

\section{Extended D-branes}
\label{extbranes}
We consider here D-branes of AdS$_3$ that are constructed from the D2-branes of the cigar.
\subsection*{H$_2$ S-brane}
These D2-branes are space-like D-branes in AdS$_3$. By descent to the coset theory 
one obtains a D2-brane of $\slc$ covering all the cigar, with a magnetic field:
\begin{equation}
F = \frac{k}{2\pi} \frac{\sin \sigma \tanh^2 r}{\sqrt{\cosh^2 r - \sin^2 \sigma}} \ 
\di r \wedge \di \phi.
\end{equation}
where $\sigma \in [0,\pi/2)$.\footnote{There are D2-branes covering only part of the cigar at the 
level of the semi-classical analysis, but their construction in the \textsc{cft} is not clear, see 
however~\cite{Fotopoulos:2004ut}.} The one point function of this D2-brane has been computed 
in~\cite{Ribault:2003ss}. However, as shown in~\cite{Israel:2004jt}, they lead to an open 
string spectrum with negative multiplicities. It was further argued that such a problem 
disappears when the level $k$ is integer~\cite{Israel:2005fn}. We will restrict our discussion to this 
case for now. The one-point function for the D2-branes then reads~\cite{Ribault:2003ss}:
\begin{eqnarray}
\langle \Phi^{j}_{nw} \rangle_{\sigma} = \frac{[k(k-2)]^{1/4}}{2\pi (k-2)}
\nu_{k}^{1/2-j}\ 
\delta_{n,0} \nonumber\\
 \Gamma \left( 2-2j  \right)
\Gamma \left( \frac{1-2j}{k-2}\right)
\frac{\Gamma \left( j+\frac{kw}{2}\right)}{\Gamma \left( 1-j+\frac{kw}{2}\right)}
   \left[ e^{i \sigma(1-2j)}  + (-)^{kw}
      e^{-i \sigma(1-2j)}  \right]
\end{eqnarray}
and the closed string annulus amplitude contains only couplings to the continuous representations. However 
it is clear that further work is needed to obtain a good understanding of this family of D-branes which 
has important applications, for instance to obtain a \textsc{cft} description of the Hanany-Witten 
effect of anomalous creation of D-branes by NS5-branes~\cite{Israel:2005fn}.

To construct the symmetric D-brane of AdS$_3$ associated with this D2-brane of the cigar we choose
Neumann boundary conditions for the extra U(1). 
We get then an Euclidean H$_2$ S-brane in Lorentzian AdS$_3$ with profile
\begin{equation}
\cosh r \sin t = \sin \sigma
\end{equation}
and magnetic field (in the first gauge) 
\begin{equation}
F = \frac{k}{2\pi} \frac{\sin \sigma \tanh r}{\sqrt{\cosh^2 r - \sin^2 \sigma}} \ \di r \wedge \di \varphi
\label{gaugeH2}
\end{equation}
more precisely, we have two copies of such an S-brane.
As for the D-instanton we can focus on the single cover of AdS$_3$ with $N=1$. 
The one-point function for the H$_2$ spacelike D-brane is then given by:
\begin{eqnarray}
\langle V^{j}_{n w \gamma p  W} \rangle = 
\frac{k\sqrt{R}}{2\pi (k-2)} \left( \frac{k-2}{2} \right)^{1/4}
\nu_{k}^{1/2-j}\ 
\delta_{n,0} \ \delta_{n+kp,0} \ 
\nonumber\\
 \Gamma \left( 2-2j  \right)
\Gamma \left( \frac{1-2j}{k-2}\right)
\frac{\Gamma \left( j+\frac{kw-\gamma}{2}\right)}{\Gamma \left( 1-j+\frac{kw-\gamma}{2}\right)}
   \left[ e^{i \sigma(1-2j)}  + (-)^{kw-\gamma}
      e^{-i \sigma(1-2j)}  \right]
\end{eqnarray}
After computations very similar to those for the AdS$_2$ D-branes, we obtain the following relative 
open string partition function:
\begin{eqnarray}
Z^\textsc{op}_{\text{H}_2} (\tau) = \sum_{n \in \zi}
\int_{0}^\infty dP \
\left\{ \frac{\partial}{2i\pi \partial P} \log \frac{R (P|i \frac{\sigma +\sigma'}{2})}{
R (P|i\frac{\sigma_\ast +\sigma_\ast '}{2})}
 \lambda^{c}_{P,n} (\tau )\right. \phantom{aaaaaaaaaaaaaaaa}
\nonumber \\ +
\left.
\frac{\partial}{2i\pi \partial P} \log \frac{R (P|i \frac{\sigma -\sigma'}{2})}{R (P|i\frac{\sigma_\ast-\sigma_\ast '}{2})}
 \lambda^{c}_{P,n+\frac{k}{2}} (\tau) \right\} \ 
\sum_{\omega \in \zi} q^{\frac{(n+k\omega )^2}{R^2 k}}
\end{eqnarray}
For the Lorentzian AdS$_3$ (i.e. $R^2 =-1$) it gives characters of continuous representations of 
$\slr$ as well as their images under spectral flow. Open strings with odd spectral flow number (second line) 
are stretched between the two copies of the H$_2$ D-brane. 
However the interpretation in spacetime 
of this open string spectrum is not clear since they are localized in the time direction.  
It should be possible to construct a boundary state describing the decay of this 
S-brane, by an analogue of the rolling tachyon solution~\cite{Callan:1994ub,Sen:2002nu}.  
\subsection*{Spacetime-filling D-brane}
If we consider instead Dirichlet boundary conditions instead for the extra U(1), we 
obtain a spacetime-filling D2-brane with a gauge field, given by the same 
expression~(\ref{gaugeH2}). the one-point function is given by:
\begin{eqnarray}
\langle V^{j}_{n w \gamma p  W} \rangle =  \frac{\sqrt{\nicefrac{k}{R}}}{2\pi^2 [2(k-2)]^{1/4}} \ \ 
\delta_{n,0} \ \delta_{NkW+\gamma,0} \ e^{iNp \, x_0} 
\nu^{\frac{1}{2}-j} \nonumber \\  \frac{\Gamma(j+\frac{kw}{2})
\Gamma(j-\frac{kw}{2})
}{\Gamma(2j-1) \Gamma(1-\frac{1-2j}{k})}
\frac{e^{i \sigma(1-2j)} \sin \pi (j-\frac{kw}{2}) +
      e^{-i \sigma(1-2j)} \sin \pi (j+\frac{kw}{2})
}{\sin \pi(1-2j) \sin \pi \frac{1-2j}{k} } 
\end{eqnarray}
The open string annulus partition function can be computed in a straightforward way, and leads to the 
expression (for the relative partition function):
\begin{eqnarray}
Z^\textsc{op}_{\text{AdS}_3} (\tau) = 
\int_{0}^\infty dP \ \sum_{n \in \zi} \ 
\left\{ \frac{\partial}{2i\pi \partial P} \log \frac{R (P|i \frac{\sigma +\sigma'}{2})}{
R (P|i\frac{\sigma_\ast +\sigma_\ast '}{2})}
 \lambda^{c}_{P,n} (\tau )\right. \phantom{aaaaaaaaaaaaaaaa}
\nonumber \\ +
\left.
\frac{\partial}{2i\pi \partial P} \log \frac{R (P|i \frac{\sigma -\sigma'}{2})}{R (P|i\frac{\sigma_\ast-\sigma_\ast '}{2})}
\lambda^{c}_{P,n+\frac{k}{2}} (\tau) \right\} \ 
\int dE \  \frac{q^{-\frac{E^2}{k}}}{\eta (\tau)}
\end{eqnarray}
giving a result very similar to the symmetry-breaking D-branes associated to the AdS$_2$ branes. 

This concludes our analysis of D-branes in AdS$_3$. We have considered all the D-branes that can be 
decomposed in the T-dual geometry in terms of D-branes in the elliptic coset $\slc$. It would be 
interesting to consider the D-branes related to the hyperbolic and parabolic cosets, for instance 
to construct D-branes in the \textsc{btz} black hole. 

\section*{Acknowledgements}
I would like to thank Costas Bachas, Shmuel Elitzur, Costas Kounnas, Ari Pakman, 
Amit Sever and Jan Troost for stimulating discussions.

\appendix

\section{Characters of $\slr$ and $\slc$}
In this section we recall some properties of the characters of $\slr$ and their 
decomposition in terms of coset characters~\cite{Bakas:1991fs} (see also~\cite{Itoh:1993mt}). 
We follow closely the presentation of~\cite{Pakman:2003kh}. We discuss the {\it elliptic basis} 
diagonalizing the action of the timelike current $J^3$. Indeed the cigar \textsc{cft} 
$\slc$ is obtained by gauging the corresponding subgroup of $\slr$.
\subsection*{Continuous representations}
The continuous representations of $\slr$ are labeled by a continuous spin $j = \nicefrac{1}{2} 
+ i P$ with $P \in \mathbb{R}_+$, and a second label $\alpha$ with $2\alpha \in \zi_{2}$ for the single cover 
of $\slr$.\footnote{Otherwise $2N\alpha \in \zi_{2N}$ for the N-th cover.}
The characters are just obtained by the free action of the modes of the currents as:
\begin{equation}
\chi^{c}_{\nicefrac{1}{2}+iP,\alpha} (\tau, \nu) = 
\text{Tr}_{\text{rep}(\nicefrac{1}{2}+iP,\alpha )} \ \left( q^{L_0 - c/24} e^{2i\pi \nu J^{3}_0 }\right) \  
=\ \frac{q^{\frac{P^2}{k-2}}}{\eta^3}\sum_{n \in \zi} e^{2i\pi \nu (\alpha+ n)}
\end{equation}
The continuous characters of the coset are obtained by decomposing the $\slr$ characters according to the 
$\widehat{\mathfrak{u}(1)}$ affine subalgebra of $\widehat{\mathfrak{sl}(2,\mathbb{R})}$ that is gauged:
\begin{equation}
\chi^{c}_{\nicefrac{1}{2}+iP,\alpha} (\tau, \nu) = 
\sum_{r \in \zi} \lambda^{c}_{\nicefrac{1}{2}+iP,r+\alpha} \ 
e^{2i\pi \nu (r+\alpha)} \frac{q^{-\frac{(r+\alpha)^2}{k}}}{\eta  (\tau )} = 
\sum_{r \in \zi} q^{\frac{P^2}{k-2}} \frac{q^{\frac{(r+\alpha)^2}{k}}}{\eta^2}
 \
e^{2i\pi (r+\alpha)} \frac{q^{-\frac{(r+\alpha)^2}{k}}}{\eta  (\tau )} 
\label{decompcont}
\end{equation}
thus giving the coset characters $\lambda^{c}_{\nicefrac{1}{2}+iP,r+\alpha}$.
Their modular transformation is straightforward and reads:
\begin{equation}
\lambda^{c}_{\nicefrac{1}{2}+iP,r+\alpha} (-1/\tau )  = \frac{4}{\sqrt{k(k-2)}} \int \di \mu \ 
e^{-4i\pi \frac{(r+\alpha) \mu}{k}} \int_{0}^{\infty} \di P' \ \cos \left( \frac{4\pi P P'}{k-2}\right) 
\lambda^{c}_{\nicefrac{1}{2}+iP',\mu} (\tau)
\end{equation}
New characters of $\slr$ are obtained by the {\it spectral flow} external automorphism of the $\widehat{\mathfrak{sl}(2,\mathbb{R})}$ 
affine algebra~\cite{Maldacena:2000hw} of integer parameter $w$:
\begin{equation}
\tilde{J}^{3}_n = J^{3}_n - \frac{kw}{2} \ \delta_{n,0} \ , \ \ 
\tilde{J}^{\pm}_n = J^{\pm}_{n\pm w}
\end{equation}
This is an external automorphism of the affine algebra generating new representations whose spectrum is not bounded 
from below. Nevertheless they are necessary to obtain a consistent AdS$_3$ string theory. The continuous 
representations with spectral flow corresponds in AdS$_3$ to "long strings" that are macroscopic fundamental strings that 
expands towards the boundary of AdS$_3$ while winding $w$ times. The corresponding characters can be written 
using the coset decomposition as:
\begin{equation}
\chi^{c \, (w)}_{\nicefrac{1}{2}+iP,\alpha} (\tau, \nu) = 
\sum_{r \in \zi} \lambda^{c}_{\nicefrac{1}{2}+iP,r+\alpha} \ 
e^{2i\pi (r+\alpha+kw/2)\nu} \frac{q^{-\frac{(r+\alpha+kw/2)^2}{k}}}{\eta  (\tau )}
\end{equation}
\subsection*{Discrete representations}
The lowest weight discrete representations are characterized by a real spin $2j \in \nicefrac{\zi}{N}$ (for the 
$N$-th cover) and the charge of the states in a representation $j$ 
under the elliptic subalgebra is $J_{0}^{3} = j +r$, with $r \in \mathbb{Z}$ 
(the primary states have $r \in \mathbb{N}$). Their characters are more involved and reads:
\begin{eqnarray}
\chi^{d}_{j} (\tau,\nu ) = \frac{q^{-\frac{(j-\nicefrac{1}{2})^2}{k-2}}\ e^{2i\pi \nu (j-\nicefrac{1}{2})}}{
i\vartheta_1 (\tau,\nu)} \nonumber \\ \text{with} \quad 
\vartheta_1 (\tau,\nu) = -i e^{-i\pi \nu} q^{3/24} \prod_{n=1}^{\infty} (1-q^n)(1-e^{2i\pi \nu}\, q^{n-1})(1-e^{-2i\pi \nu}\, q^n)
\end{eqnarray}
Then the characters of the coset are obtained by a decomposition similar to~(\ref{decompcont}), and reads:
\begin{equation}
\lambda^{d}_{j,r} (\tau ) = \frac{q^{-\frac{(j-\nicefrac{1}{2})^2}{k-2}}}{\eta^2 (\tau )}
\ S_r (\tau)\  q^{\frac{(j+r)^2}{k}} \quad \text{with} \quad 
S_r (\tau ) = \sum_{n=0}^\infty (-)^n q^{\frac{n(n+2r+1)}{2}}
\end{equation}
The modular transformation of these characters are not known. However, for integer $k$ one can define 
{\it extended characters} as $\Lambda^{d}_{j,n} = \sum_{\omega} \lambda^{d}_{j,n+k\omega}$. Their modular transformation 
has been computed in~\cite{Israel:2004xj}.

For the discrete representations one can also define spectral flowed characters of $\slr$, 
that can be written as:
\begin{equation}
\chi^{d \, (w)}_{j} (\tau, \nu) = 
\sum_{r \in \zi} \lambda^{j}_{j,r} \ 
e^{2i\pi (j+r+kw/2)\nu} \frac{q^{-\frac{(j+r+kw/2)^2}{k}}}{\eta  (\tau )}
\end{equation}
\subsection*{Identity representation}
The character of $\slr$ associated to the identity representation reads:
\begin{equation}
\chi^\mathbb{I} (\tau , \nu)  = -\frac{q^{-\frac{1}{4(k-2)}}\ 2\sin \pi \nu}{\vartheta_1 (\tau, \nu)}
\end{equation}
and the only primary state of the affine algebra has $J^{3}_0 = 0$. 
It is expanded in terms of coset characters as:
\begin{equation}
\chi^\mathbb{I} (\tau , \nu) = \sum_{r\in \zi} \lambda^{\mathbb{I}}_r (\tau) 
\ e^{2i\pi r} \frac{q^{-\frac{r^2}{k}}}{\eta  (\tau )}
\end{equation}
where
\begin{equation}
\lambda^{\mathbb{I}}_r (\tau)  = 
\frac{q^{-\frac{1}{4(k-2)}}}{\eta (\tau)^2 } q^{|r|+\frac{r^2}{k}}  \left[
1+\sum_{n=1}^{\infty} (-)^n q^{\frac{n^2+(2|r|+1)n-2|r|}{2}} (1+q^{|r|})\right].
\end{equation}
One can define spectral flowed identity characters as well.

\subsection*{Fourier transform}
The bulk two-point function in the coset $\slc$ has been obtained in~\cite{Giveon:1999tq} by using the following Fourier transform:
\begin{equation}
\Phi^{j}_{m\bar{m}} = \frac{1}{4\pi^2} \int_\mathbb{C} d^2 x\  x^{j+m-1} \bar{x}^{j+\bar{m}-1}\  \Phi^{j} (x,\bar{x})
\end{equation}
in order to diagonalize the action of $J^{3}$. The fields in H$_3^+$ satisfy the following 
reflection property~\cite{Teschner:1997ft}:
\begin{equation}
\Phi^{j} (x,\bar{x}) =  R(j) \frac{1-2j}{\pi} \int d^2 u \ |u-x|^{-4j} \ \Phi^{1-j} (u,\bar{u})
\end{equation}
written in terms of the \textsc{cft} reflection amplitude 
\begin{equation}
R(j) = - \nu^{1-2j} \frac{\Gamma (1+\frac{1-2j}{k-2})}{\Gamma (1-\frac{1-2j}{k-2})}
\end{equation}
and involving the intertwining operator $\mathcal{I}^j$ realizing the isomorphism of the representations 
of spin $j$ and $1-j$ in SL(2,$\mathbb{C}$), of kernel:
\begin{equation}
I^j (u|x) = \frac{1-2j}{\pi} |u-x|^{-4j}
\end{equation}
normalized such that $\mathcal{I}^{1-j} \circ \mathcal{I}^j = \mathbb{I}$.
Then one obtains by the Fourier transform the reflection amplitude in the coset that is used to write 
eq.~(\ref{2ptfctbulk}).

Similarlily one can define a Fourier transform for the boundary operators corresponding to AdS$_2$ branes 
with boundary conditions $(r \, r')$ by:
\begin{equation}
\Phi^{j \ (rr')}_m  = \int_\mathbb{R} dx \ x^{j-1+m} \ \Phi^{j\ (rr')} (x) 
\end{equation}
and in this basis these are eigenfunctions of $J^3$ with eigenvalue $m$. 
The reflection property for such boundary fields reads~\cite{Ponsot:2001gt,Ribault:2002ti}:
\begin{equation}
\Phi^{j\ (rr')} (x) = R(j|rr')\ 
(\mathcal{J}^j \Phi^{1-j\ (rr')} ) (u)
\end{equation}
where the boundary reflection amplitude is:
\begin{eqnarray}
R(j|rr') = \nu^{\nicefrac{1}{2}-j} \ \frac{\Gamma^{2}_k (k-2+j)}{\Gamma^{2}_k (k-2+1-j)}
\frac{\Gamma_k ( k-2 +1-2j)}{\Gamma_k ( k-2 +2j-1)} \nonumber\\
\frac{S_{k}^{(0)} \left(\frac{k-2}{2\pi} (r+r') -i(j-1/2) \right)}{
S_{k}^{(0)} \left(\frac{k-2}{2\pi} (r+r') +i(j-1/2) \right)}
\frac{S_{k}^{(1)} \left(\frac{k-2}{2\pi} (r-r') -i(j-1/2) \right)}{
S_{k}^{(1)} \left(\frac{k-2}{2\pi} (r-r') +i(j-1/2) \right)}
\frac{S_{k}^{(0)}(+i(j-1/2))}{S_{k}^{(0)}(-i(j-1/2))}\nonumber\\
\end{eqnarray}
and the intertwining operator $\mathcal{J}^j$ realizing the isomorphism of the representations 
of spin $j$ and $1-j$ in SL(2,$\mathbb{R}$) is also normalized such that 
$\mathcal{J}^{1-j} \circ \mathcal{J}^j = \mathbb{I}$:
\begin{equation}
( \mathcal{J}^j \Phi^{1-j\ (rr')}  ) (u)  = \frac{1}{\Gamma (1-2j) (1+e^{i\pi (2j-1)})} 
\int_\mathbb{R} dv \ |u-v|^{-2j} \ \Phi^{1-j\ (rr')} (v)
\end{equation}

Consequently after a Fourier transform it gives the reflection amplitude given in the text, 
eq.~(\ref{boundreflampl}).


\begin{thebibliography}{99}

\bibitem{Maldacena:1997re}
J.~M.~Maldacena,
``The large N limit of superconformal field theories and supergravity,''
Adv.\ Theor.\ Math.\ Phys.\  {\bf 2} (1998) 231
[Int.\ J.\ Theor.\ Phys.\  {\bf 38} (1999) 1113]
[arXiv:hep-th/9711200].

\bibitem{Banados:1992wn}
M.~Banados, C.~Teitelboim and J.~Zanelli,
``The Black hole in three-dimensional space-time,''
Phys.\ Rev.\ Lett.\  {\bf 69} (1992) 1849
[arXiv:hep-th/9204099].


\bibitem{Maldacena:2000hw}
J.~M.~Maldacena and H.~Ooguri,
``Strings in AdS(3) and SL(2,R) WZW model. I,''
J.\ Math.\ Phys.\  {\bf 42}, 2929 (2001)
[arXiv:hep-th/0001053].


\bibitem{Henningson:1991jc}
M.~Henningson, S.~Hwang, P.~Roberts and B.~Sundborg,
``Modular invariance of SU(1,1) strings,''
Phys.\ Lett.\ B {\bf 267}, 350 (1991).

\bibitem{Israel:2003ry}
D.~Israel, C.~Kounnas and M.~P.~Petropoulos,
``Superstrings on NS5 backgrounds, deformed AdS(3) and holography,''
JHEP {\bf 0310} (2003) 028
[arXiv:hep-th/0306053].


\bibitem{Teschner:1997ft}
J.~Teschner,
``On structure constants and fusion rules in the SL(2,C)/SU(2) WZNW  model,''
Nucl.\ Phys.\ B {\bf 546}, 390 (1999)
[arXiv:hep-th/9712256].

\bibitem{Teschner:1997fv}
J.~Teschner,
``The mini-superspace limit of the SL(2,C)/SU(2) WZNW model,''
Nucl.\ Phys.\ B {\bf 546}, 369 (1999)
[arXiv:hep-th/9712258].


\bibitem{Teschner:1999ug}
J.~Teschner,
``Operator product expansion and factorization in the H-3+ WZNW model,''
Nucl.\ Phys.\ B {\bf 571}, 555 (2000)
[arXiv:hep-th/9906215].

\bibitem{Teschner:2001gi}
J.~Teschner,
``Crossing symmetry in the H(3)+ WZNW model,''
Phys.\ Lett.\ B {\bf 521}, 127 (2001)
[arXiv:hep-th/0108121].


\bibitem{Maldacena:2001km}
J.~M.~Maldacena and H.~Ooguri,
``Strings in AdS(3) and the SL(2,R) WZW model. III: Correlation  functions,''
Phys.\ Rev.\ D {\bf 65}, 106006 (2002)
[arXiv:hep-th/0111180].

\bibitem{Elitzur:1991cb}
S.~Elitzur, A.~Forge and E.~Rabinovici,
``Some global aspects of string compactifications,''
Nucl.\ Phys.\  {\bf B359}, 581 (1991).

\bibitem{Mandal:1991tz}
G.~Mandal, A.~M.~Sengupta and S.~R.~Wadia,
``Classical solutions of two-dimensional string theory,''
Mod.\ Phys.\ Lett.\  {\bf A6}, 1685 (1991).

\bibitem{Witten:1991yr}
E.~Witten,
``On string theory and black holes,''
Phys.\ Rev.\  {\bf D44} (1991) 314.


\bibitem{Bachas:2000fr}
C.~Bachas and M.~Petropoulos,
``Anti-de-Sitter D-branes,''
JHEP {\bf 0102}, 025 (2001)
[arXiv:hep-th/0012234].

\bibitem{Petropoulos:2001qu}
P.~M.~Petropoulos and S.~Ribault,
``Some remarks on anti-de Sitter D-branes,''
JHEP {\bf 0107}, 036 (2001)
[arXiv:hep-th/0105252].


\bibitem{Hikida:2001yi}
  Y.~Hikida and Y.~Sugawara,
  ``Boundary states of D-branes in AdS(3) based on discrete series,''
  Prog.\ Theor.\ Phys.\  {\bf 107} (2002) 1245
  [arXiv:hep-th/0107189].

\bibitem{Rajaraman:2001cr}
  A.~Rajaraman and M.~Rozali,
  ``Boundary states for D-branes in AdS(3),''
  Phys.\ Rev.\ D {\bf 66} (2002) 026006
  [arXiv:hep-th/0108001].

\bibitem{Deliduman:2002bf}
  C.~Deliduman,
  ``AdS(2) D-branes in Lorentzian AdS(3),''
  Phys.\ Rev.\ D {\bf 68} (2003) 066006
  [arXiv:hep-th/0211288].







\bibitem{Lee:2001xe}
P.~Lee, H.~Ooguri, J.~W.~Park and J.~Tannenhauser,
``Open strings on AdS(2) branes,''
Nucl.\ Phys.\  {\bf B610}, 3 (2001)
[arXiv:hep-th/0106129].


\bibitem{Ponsot:2001gt}
B.~Ponsot, V.~Schomerus and J.~Teschner,
``Branes in the Euclidean AdS(3),''
JHEP {\bf 0202}, 016 (2002)
[arXiv: hep-th/0112198].

\bibitem{Lee:2001gh}
P.~Lee, H.~Ooguri and J.~w.~Park,
``Boundary states for AdS(2) branes in AdS(3),''
Nucl.\ Phys.\  {\bf B632}, 283 (2002)
[arXiv:hep-th/0112188].

\bibitem{Ribault:2003ss}
S.~Ribault and V.~Schomerus,
``Branes in the 2-D black hole,''
JHEP {\bf 0402}, 019 (2004)
[arXiv:hep-th/0310024].

\bibitem{Israel:2004jt}
D.~Israel, A.~Pakman and J.~Troost,
``D-branes in N = 2 Liouville theory and its mirror,''
arXiv: hep-th/0405259.

\bibitem{Eguchi:2003ik}
T.~Eguchi and Y.~Sugawara,
``Modular bootstrap for boundary N = 2 Liouville theory,''
JHEP {\bf 0401}, 025 (2004)
[arXiv:hep-th/0311141].

\bibitem{Fotopoulos:2004ut}
A.~Fotopoulos, V.~Niarchos and N.~Prezas,
``D-branes and extended characters in SL(2,R)/U(1),''
arXiv:hep-th/0406017.


\bibitem{Maldacena:2001ky}
J.~M.~Maldacena, G.~W.~Moore and N.~Seiberg,
``Geometrical interpretation of D-branes in gauged WZW models,''
JHEP {\bf 0107} (2001) 046
[arXiv: hep-th/0105038].




\bibitem{Sfetsos:1998xd}
K.~Sfetsos,
``Branes for Higgs phases and exact conformal field theories,''
JHEP {\bf 9901}, 015 (1999)
[arXiv:hep-th/9811167].


\bibitem{Israel:2004ir}
D.~Israel, C.~Kounnas, A.~Pakman and J.~Troost,
``The partition function of the supersymmetric two-dimensional black hole and
little string theory,''
JHEP {\bf 0406}, 033 (2004)
[arXiv:hep-th/0403237].

\bibitem{Ooguri:1995wj}
H.~Ooguri and C.~Vafa,
``Two-Dimensional Black Hole and Singularities of CY Manifolds,''
Nucl.\ Phys.\ B {\bf 463}, 55 (1996)
[arXiv:hep-th/9511164].

\bibitem{Giveon:1999px}
A.~Giveon and D.~Kutasov,
``Little string theory in a double scaling limit,''
JHEP {\bf 9910}, 034 (1999)
[arXiv:hep-th/9909110].




\bibitem{Giveon:1993ph}
A.~Giveon and E.~Kiritsis,
``Axial vector duality as a gauge symmetry and topology change in string
theory,''
Nucl.\ Phys.\  {\bf B411} (1994) 487
[arXiv:hep-th/9303016].

\bibitem{Israel:2005fn}
D.~Israel, A.~Pakman and J.~Troost,
``D-branes in little string theory,''
arXiv:hep-th/0502073.



\bibitem{Hanany:2002ev}
A.~Hanany, N.~Prezas and J.~Troost,
``The partition function of the two-dimensional black hole conformal  field
theory,''
JHEP {\bf 0204}, 014 (2002)
[arXiv:hep-th/0202129].







\bibitem{Maldacena:2000kv}
J.~M.~Maldacena, H.~Ooguri and J.~Son,
``Strings in AdS(3) and the SL(2,R) WZW model. II: Euclidean black hole,''
J.\ Math.\ Phys.\  {\bf 42} (2001) 2961
[arXiv:hep-th/0005183].

\bibitem{Giveon:1999tq}
A.~Giveon and D.~Kutasov,
``Comments on double scaled little string theory,''
JHEP {\bf 0001} (2000) 023
[arXiv:hep-th/9911039].

\bibitem{Bachas:2001vj}
C.~Bachas, J.~de Boer, R.~Dijkgraaf and H.~Ooguri,
``Permeable conformal walls and holography,''
JHEP {\bf 0206} (2002) 027
[arXiv:hep-th/0111210].

\bibitem{Aharony:2004xn}
O.~Aharony, A.~Giveon and D.~Kutasov,
``LSZ in LST,''
Nucl.\ Phys.\  {\bf B691}, 3 (2004)
[arXiv:hep-th/0404016].

\bibitem{Simon:1998az}
J.~Simon,
``T-duality and effective D-brane actions,''
arXiv:hep-th/9812095.


\bibitem{Cardy:1989ir}
J.~L.~Cardy,
``Boundary Conditions, Fusion Rules And The Verlinde Formula,''
Nucl.\ Phys.\  {\bf B324} (1989) 581.


\bibitem{Ribault:2002ti}
S.~Ribault,
``Two AdS(2) branes in the Euclidean AdS(3),''
JHEP {\bf 0305}, 003 (2003)
[arXiv:hep-th/0210248].

\bibitem{Sarkissian:2002ie}
G.~Sarkissian,
``Non-maximally symmetric D-branes on group manifold in the Lagrangian
approach,''
JHEP {\bf 0207} (2002) 033
[arXiv:hep-th/0205097].


\bibitem{Quella:2002ns}
T.~Quella,
``On the hierarchy of symmetry breaking D-branes in group manifolds,''
JHEP {\bf 0212} (2002) 009
[arXiv:hep-th/0209157].

\bibitem{Callan:1994ub}
C.~G.~.~Callan, I.~R.~Klebanov, A.~W.~W.~Ludwig and J.~M.~Maldacena,
``Exact solution of a boundary conformal field theory,''
Nucl.\ Phys.\  {\bf B422} (1994) 417
[arXiv:hep-th/9402113].

\bibitem{Sen:2002nu}
A.~Sen,
``Rolling tachyon,''
JHEP {\bf 0204}, 048 (2002)
[arXiv:hep-th/0203211].

\bibitem{Bakas:1991fs}
I.~Bakas and E.~Kiritsis,
``Beyond the large N limit: Nonlinear W(infinity) as symmetry of the SL(2,R) /
U(1) coset model,''
Int.\ J.\ Mod.\ Phys.\  {\bf A7}, 55 (1992)
[arXiv:hep-th/9109029].

\bibitem{Itoh:1993mt}
  K.~Itoh, H.~Kunitomo, N.~Ohta and M.~Sakaguchi,
  ``BRST Analysis of physical states in two-dimensional black hole,''
  Phys.\ Rev.\ D {\bf 48} (1993) 3793
  [arXiv:hep-th/9305179].




\bibitem{Pakman:2003kh}
A.~Pakman,
``BRST quantization of string theory in AdS(3),''
JHEP {\bf 0306}, 053 (2003)
[arXiv:hep-th/0304230].

\bibitem{Israel:2004xj}
D.~Israel, A.~Pakman and J.~Troost,
``Extended SL(2,R)/U(1) characters, or modular properties of a simple
non-rational conformal field theory,''
JHEP {\bf 0404} (2004) 045
[arXiv:hep-th/0402085].


\end{thebibliography}
\end{document}